\documentstyle[epsf,aps,12pt]{revtex}
\newcommand{\beq}{\begin{equation}}
\newcommand{\eeq}{\end{equation}}
\newcommand{\beqa}{\begin{eqnarray}}
\newcommand{\eeqa}{\end{eqnarray}}

\tighten

\title{Pseudogap in 1d revisited}
\author{Oleg Tchernyshyov\thanks{
Current address: School of Natural Sciences, Institute for Advanced Study,
Princeton, NJ 08540.}}

\address{Physics Department, Columbia University, New York, NY 10027}

\begin{document}
\maketitle

\begin{abstract}
Two decades ago, Sadovskii found an exact solution of a model
describing a pseudogap in electron energy spectrum 
(first introduced by Lee, Rice and Anderson).  
The discovery of a pseudogap in high-$T_c$ superconductors has
revived the interest to his exact solution.  I review the model with 
the emphasis on physical content, point out an error in the original 
Sadovskii's solution and explain which problem he actually solved.  
A recent incorporation of Sadovskii's ideas into a description of 
``hot spots'' on the Fermi surface in cuprate superconductors 
(Schmalian, Pines and  Stojkovi\'c) is briefly discussed.
\end{abstract}


\section{Introduction}

A model of electrons with a pseudogap from fluctuations
of an order parameter was introduced in 1973 by P.~A.~Lee, T.~M.~Rice 
and P.~W.~Anderson \cite{Lee73}.  
A few years later, M.~V.~Sadovskii showed that it admits an exact 
solution \cite{Sad73,Sad79}.  
The model describes a Peierls system (a metallic chain with a charge 
density wave instability) above the phase transition temperature $T_P$.  
The exact solvability comes at a price: 
(a) The solution is specifically tailored for one dimension.  
(b) It is assumed that Peierls-Kohn phonons are 
described by a non-selfinteracting boson field.  These two 
limitations of the Sadovskii's solution have been known since 
its publication.  

Recently, however, I discovered an unfortunate 
error in the original paper by Sadovskii and now I am convinced that he
actually solved a completely different, rather unphysical problem.  
This and the fact that Sadovskii's work is often regarded as the
one and only exact model of the pseudogap \cite{McKenzie,Schmalian}
has prompted me to review this model.  While its mathematical side
has been discussed quite thoroughly by Sadovskii himself, 
the physical content deserves further comment.  

The plan of the paper is as follows.  After a brief description
of the Peierls instability in a one-dimensional conductor 
(Sec.~\ref{Peierls described}), a suitable mathematical formalism 
will be presented in Sec.~\ref{continuum form}.  It will be
shown that assessment of multi-phonon contributions 
to the fermion energy spectrum requires a knowledge of
statistical properties of the phonon ensemble.  The model of Sadovskii,
which postulates Gaussian statistics for the phonons, is introduced, 
interpreted and thoroughly illustrated in Sec.~\ref{Sadovskii's model}.  
This is done in order to demystify its well-known yet strange-looking
electron spectrum in the limit of long-range phonon correlation length $\xi$.  
I will then point out a previously unnoticed error in Sadovskii's 
``exact'' solution for a finite $\xi$ (Sec.~\ref{Sadovskii wrong}) 
and explain which problem Sadovskii has actually solved.  Finally,
a recent extension of the Sadovskii model to ``higher'' dimensions 
by Schmalian et al.~\cite{Schmalian} in the context of high-$T_c$ 
superconductivity will be discussed in Sec.~\ref{altius}.  

\subsection{Peierls instability and fluctuations in 1 dimension}
\label{Peierls described}

An excellent introduction to the Peierls effect can be found in
G.~Gr\"uner's book \cite{Gruner}.  See also an article by 
G.~A.~Toombs \cite{Toombs}, which reviews in detail theoretical 
and experimental developments prior to 1984.  
In a one-dimensional electron gas (with Fermi momentum $p_F$)
fluctuations of electron density are particularly strong near the 
wavevectors $\pm 2 p_F$.  This happens because 
creating a hole and an electron with momenta near $\pm p_F$ 
costs little energy.  Therefore, when electron-phonon interaction 
couples lattice vibrations to these fluctuations, phonon modes 
with momenta near $2p_F$ become ``soft'' and a static charge-density wave
(CDW) appears below a transition temperature $T_P$.  

\begin{figure}
\begin{center}
\leavevmode
\epsfxsize 2in
\epsffile{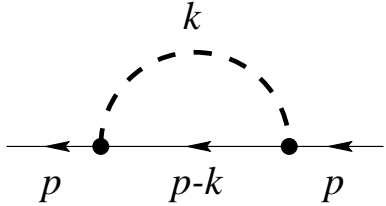}
\end{center}
\caption{One-phonon correction to the fermion propagator, 
A phonon with momentum $2p_F + k$ (dashed line) connects
electron states with momenta $p_F + p$  and $-p_F + p - k$, 
respectively energies $pv$ and $-(p-k)v$.}
\label{LRA}
\end{figure}

In the context of the mean-field theory of a Peierls transition, 
an energy gap opens at the Fermi points exactly at $T=T_P$.  However, 
some remnant of the
gap can be created by fluctuations even above $T_P$.  Lee, Rice 
and Anderson calculated the electron self-energy induced by the 
emission and reabsorption of a (dressed) phonon (Fig.~\ref{LRA}).
Approximating the two-point phonon correlation by two Lorentzian lines
peaked at $\pm 2p_F$, one obtains \cite{Lee73}
\beq
\Sigma(\omega, p_F+p) 
= \frac{\delta^2}{\omega+pv+iv\xi^{-1}},
\label{LRA Sigma}
\eeq
where 
$\delta^2 = 4\pi^3 T_P^2\xi/7\zeta(3)v$
can be regarded as the average fluctuation of the order parameter, 
$\xi$ is the phonon correlation length and $v$ the Fermi velocity.

\subsection{Continuum formulation}
\label{continuum form}

In the vicinity of the two Fermi points, the equations of motion
for the fermion field $c_k(t)$ in the presence of longitudinal phonons is 
\beqa
(id/dt - vq)c_{p_F+q}(t) &=& gL^{-1/2}\int\frac{dp}{2\pi}\ 
\delta x_{2p_F+q-p}\ c_{-p_F+p}(t),\nonumber \\
(id/dt + vq)c_{-p_F+q}(t) &=& g^*L^{-1/2}\int\frac{dp}{2\pi}\ 
\delta x_{-2p_F+q-p}\ c_{p_F+p}(t).
\label{near p_F}
\eeqa
Here $\delta x$ is the atomic displacement along the chain direction, 
$g$ is the electron-phonon coupling and $L$ is the chain length.   
The electron energy spectrum near $\pm p_F$ has been linearized,
$\epsilon_{\pm p_F+q}\approx \pm vq$.  

It is convenient to combine 
right and left-moving fermion fields into a column
\beq
\psi_q = 
\left(\begin{array}{l}
\psi_{Rq}\\
\psi_{Lq}
\end{array}\right)
= 
\left(\begin{array}{l}
c_{p_F+q}\\
c_{-p_F+q}
\end{array}\right).
\eeq
Phonons can be described as a complex gap field $\Delta(t,x)$ defined in 
terms of the Fourier transform of the displacement
\beq
\Delta_k(t) = g\, \delta x_{2p_F+k}(t), 
\hskip 1cm
{\Delta\!^*}_k(t) = g^*\, \delta x^*_{2p_F+k}(t) = g^* x_{-2p_F-k}(t).
\label{Delta defined}
\eeq
The last equality is a statement that atomic diaplacements are real.  
In contrast, 
$\Delta\!^*\,_{k}\neq\Delta_{-k}$, whereby $\Delta\!^*(t,x)\neq\Delta(t,x)$, 
i.e., the gap field $\Delta(t,x)$ is genuinely complex, 
except when the CDW is commensurate with the lattice, $2p_F = \pi/a$ or 
$2\pi/a$.

In the new notation, Eqs.~(\ref{near p_F}) can be written as
\beqa
i(\partial/\partial t + v\,\partial/\partial x)\,\psi_R(t,x)
&=& \Delta(t,x)\, \psi_L(t,x),\nonumber \\
i(\partial/\partial t - v\,\partial/\partial x)\,\psi_L(t,x)
&=& \Delta\!^*(t,x)\,\psi_R(t,x).
\eeqa
In what follows, units in which $\hbar=v=1$ will often be employed 
to simplify the notation.    

The fermion propagator for the ground state $|0\rangle$ can be defined as 
a $2\!\times\!2$ matrix $\hat{G}$ with matrix elements
\beq
G_{\sigma\sigma'}(t-t',x-x') 
= -i\langle0|
T[\psi_\sigma(t,x)\psi^\dagger_{\sigma'}(t',x')]
|0\rangle,
\eeq
$\sigma=1$ for right and $-1$ for left fermions. 
Thermal Green's functions can be defined in a similar way. 
The propagator matrix satisfies the equation
\beq
\left[
i\frac{\partial}{\partial t} + i\hat{\sigma}_3\, \frac{\partial}{\partial x}
- \hat{\Delta}(t,x)
\right]\hat{G}(t-t',x-x') 
= \delta(t-t')\delta(x-x'), 
\label{exact G hat}
\eeq
where $\hat{\Delta}(t,x)$ is the off-diagonal matrix 
$\Delta(t,x)\hat{\sigma}_+ + \Delta\!^*(t,x)\hat{\sigma}_-$ and 
$\hat{\sigma}_i$ are the Pauli matrices.  

The free [$\Delta(t,x)=0$] propagator $\hat{G}^{(0)}$ is diagonal in the
basis of left and right-moving fermions, where $\hat{\sigma}_3=\sigma=\pm1$:
\beq
G^{(0)}_{\sigma\sigma}(t,x) = -\frac{1/2\pi}{vt-\sigma x-i0\,\mbox{sign}(t)}.
\label{G0(t,x)}
\eeq
We will use extensively its Fourier transforms, 
\beqa
G^{(0)}_{\sigma\sigma}(\omega,p) 
&=& \frac{1}{\omega-\sigma p + i0\,\mbox{sign}(\omega)},
\label{G0(omega,p)}\\
G^{(0)}_{\sigma\sigma}(\omega,x) 
&=& -i\, \mbox{sign}(\omega)\, \theta(\sigma\omega x)\, e^{i\sigma\omega x},
\label{G0(omega,x)}
\eeqa
where $\theta(x)$ is the unit step-function.  Eq.~(\ref{G0(omega,x)}) 
indicates that fermions can only propagate in a single direction.  
Unless specified otherwise, it will be assumed throughout the paper 
that $\omega>0$.

The gap field $\Delta(t,x)$ is considered to be static, $\Delta(x)$.  
In thermal field theory, this corresponds to a classical 
approximation, in which the typical frequency of a boson is much
less than the temperature (and the occupation number of that mode 
greatly exceeds 1).  As long as this does not lead to an ultraviolet
catastrophe, it appears to be a reasonable approximation.  

All we need now to determine the properties of the fermions are the
correlation functions for the gap field.  In the symmetric phase 
(above $T_P$), 
\beq
\langle\Delta(x)\rangle = 
\langle\Delta\!^*(x)\rangle = 0,
\hskip 1cm
\langle\Delta(x)\Delta(x')\rangle = 
\langle\Delta\!^*(x)\Delta\!^*(x')\rangle = 0.
\label{means}
\eeq
The two-point correlation function and its Fourier transform are
\beqa
D(x-x') &\equiv& \langle\Delta(x)\Delta\!^*(x')\rangle = 
\delta^2 e^{-|x-x'|/\xi}
\label{exponential correlation}\\
D(k) &=& \delta^2\frac{2\xi^{-1}}{k^2+\xi^2}.
\label{Lorentzian line}
\eeqa

\subsection{Fermion spectrum to order $\delta^2$}
\label{G to 2nd order}

The free ($\Delta=0$) fermion density of states ${\cal N}^{(0)}(\omega)$ 
can be read off directly from the propagator $\hat{G}^{(0)}(\omega,x)$:
\beq
{\cal N}^{(0)}(\omega) = -\frac{1}{\pi}\ 
\mbox{Tr Im\,}\hat{G}^{(0)}(\omega,0)
\equiv -\frac{1}{\pi}\sum_{\sigma=\pm 1}
\left.\mbox{Im}G_{\sigma\sigma}^{(0)}(\omega,x)\right|_{x=0}.
\eeq
While the value of the Green's function (\ref{G0(omega,x)}) is not
defined at $x=0$, we can either take the limit $x\to 0^+$ or integrate
over momenta 
\beq
-\pi^{-1}\mbox{Im\,}G_{\sigma\sigma}^{(0)}(\omega,p) 
= \delta(\omega-\sigma p).
\eeq
Either way, the free density of states (per spin) is 
\beq
{\cal N}^{(0)}(\omega) = 1/\pi = 1/\pi v,
\eeq
as one expects in one dimension.  

The fermion Green's function in the presence of a gap field can be 
obtained by starting with the free propagator and iterating
Eq.~(\ref{exact G hat}).  This procedure gives an expansion of 
$\hat{G}$ in powers of the gap field,
\beq
\langle\hat{G}\rangle = \hat{G}^{(0)} + \langle\hat{G}^{(2)}\rangle + 
\langle\hat{G}^{(4)}\rangle +\ldots
\eeq
The brackets signify averaging over configurations of the phonon field.  
The lowest-order correction $\langle \hat{G}^{(2)}\rangle$
is a diagonal matrix.  E.g., for right-moving fermions, 
\beq
G_{RR}^{(2)}(x',x) = \int d\zeta\, d\zeta'\ 
G_{RR}^{(0)}(x'-\zeta)\, \Delta(\zeta)\, G_{LL}^{(0)}(\zeta-\zeta')\,
\Delta\!^*(\zeta')\, G_{RR}^{(0)}(\zeta'-x).
\eeq
Averaging over a phonon ensemble with a mean fluctuation 
(\ref{exponential correlation})
brings out 
$\langle\Delta(\zeta)\, \Delta\!^*(\zeta')\rangle 
= \delta^2e^{|\zeta-\zeta'|/\xi}$.  Then 
\beq
\langle G_{RR}^{(2)}(x',x) \rangle = \delta^2 \int d\zeta\, d\zeta'\ 
e^{-|\zeta-\zeta'|/\xi}\,
G_{RR}^{(0)}(x'-\zeta)\, 
G_{LL}^{(0)}(\zeta-\zeta')\,
G_{RR}^{(0)}(\zeta'-x),
\eeq
which is translationally invariant.  

Surely this correction can be computed in an easier way, by working 
directly with Fourier transformed quantities, 
Fig.~\ref{LRA} and Eq.~(\ref{LRA Sigma}).  
Coordinate representation, neverthelss, is also useful.  
After all, the local density of states is given by 
$-\pi^{-1}\mbox{Tr Im\,}\hat{G}(\omega;x,x)$.  
As a bonus, we will see where and why Sadovskii's exact solution 
actually works --- see Sec.~\ref{Which model}.

\begin{figure}
\begin{center}
\leavevmode
\epsfxsize 2.5in
\epsffile{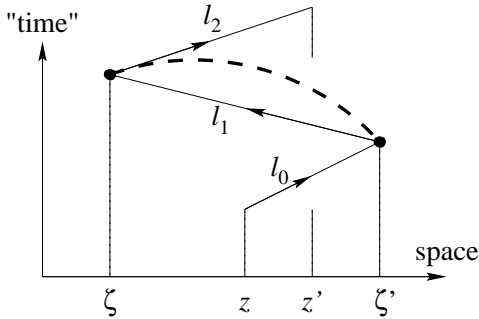}
\end{center}
\caption{Second-order correction to the fermion propagator 
$G^{(2)}(\omega,x',x)$.
Solid lines: free fermion propagator $-ie^{i\omega l_n}$.  
Dashed line: two-point phonon correlation 
$\delta^2e^{-|\zeta-\zeta'|/\xi} = \delta^2e^{-l_1/\xi}$.
The ``time'' direction is added to split apart fermion lines.
}
\label{paths2}
\end{figure}

Thanks to the presence of a step-function in the free propagators 
(\ref{G0(omega,x)}), it is more convenient to integrate over path lengths
$l_1, l_2, l_3$ than over intermediate coordinates $\zeta_1$, 
$\zeta_2$ (Fig.~\ref{paths2}).  When $\omega>0$, the free fermion propagator
is
\beq
G_{\sigma\sigma}^{(0)} = -ie^{i\omega l_n}, 
\hskip 1cm 
0 \leq l_n < \infty.  
\eeq
The lengths of fermion legs are not completely independent as the total 
displacement $x'-x$ is fixed.  This constraint is implemented by inserting
\beq
\delta(x - x' - l_0 + l_1 - l_2) 
= \int\frac{dp}{2\pi}\ e^{ip(x-x' - l_0 + l_1 - l_2)}
\eeq
in the integrand.  

In particular, when we are interested in the local density of states,
it makes sense to evaluate
\beqa
\langle G_{RR}^{(2)}(x,x)\rangle 
&=& i\delta^2 \int\frac{dp}{2\pi}
\int_0^\infty\!\!dl_2\, e^{i(\omega-p)l_2}
\int_0^\infty\!\!dl_1\, e^{i(\omega+p+i/\xi)l_1}
\int_0^\infty\!\!dl_0\, e^{i(\omega-p)l_0}
\nonumber\\
&=& \delta^2 \int\frac{dp}{2\pi}\ 
\frac{1}{\omega-p+i0}\ 
\frac{1}{\omega+p+i\xi^{-1}}\ 
\frac{1}{\omega-p+i0}.
\label{G2(x,x')}
\eeqa
The integrand on the last line is the familiar second-order
self-energy (\ref{LRA Sigma}) with the external legs reattached.  
After the integration, we find the density of states to order $\delta^2$:
\beq
{\cal N}^{(0)}(\omega) + {\cal N}^{(2)}(\omega) 
= \frac{1}{\pi v}
\left(
1 + \mbox{Re}\,\frac{2\delta^2}{(2\omega+iv\xi^{-1})^2}
\right)
\label{DOS to 2nd order}
\eeq
(the factor of 2 comes from adding the contribution of left-moving fermions).
The density of states is reduced in the range $|\omega|<v\xi^{-1}/2$, 
which can be called a pseudogap.  This approximation is valid only
when the fluctuations are fast enough, $\delta \ll v\xi^{-1}/2$.

\subsection{Beyond $\delta^2$}
\label{beyond 2nd order}

Correction to the fermion propagator of order $\delta^{2n}$ reads 
\beqa
\langle G_{RR}^{(2n)}(x',x)\rangle 
&=& \int\! d\zeta_{n}\,d\zeta_n'\ldots d\zeta_1\,d\zeta_1'\ 
G_{RR}^{(0)}(x'-\zeta_{n})\ldots 
G_{LL}^{(0)}(\zeta_1-\zeta_1')\,
G_{RR}^{(0)}(\zeta_1'-x)
\nonumber\\
&&\times D(\zeta_n,\zeta_n',\ldots,\zeta_1,\zeta_1'),
\label{G2n}
\eeqa
where the $2n$-point correlation functions is
\beq
D(\zeta_n,\zeta_n',\ldots,\zeta_1,\zeta_1') = 
\langle
\Delta(\zeta_n)\, \Delta\!^*(\zeta_n')\ldots
\Delta(\zeta_1)\, \Delta\!^*(\zeta_1')
\rangle 
\label{D2n}
\eeq
In principle, the $2n$-point correlation function must be determined 
in a microscopic theory.  In most cases, however, evaluation of 
higher-order phonon correlation functions 
is a rather difficult job.  Alternatively, one can try to see what 
comes out of (\ref{G2n}) given a certain statistics of the 
phonon field (\ref{D2n}).   

A rather trivial example would be that of the mean-field approximation,
in which the displacement amplitude is uniform throughout the chain,
i.e., $\Delta(\zeta) = \delta$ with certainty for any 
$\zeta$.  In this case, (\ref{D2n}) reduces to 
\beq
D(\zeta_n,\zeta_n',\ldots,\zeta_1,\zeta_1') = 
\delta^{2n}.
\label{MF statistics}
\eeq
Then, the Fourier transform of the right-moving propagator is
\beqa
\langle G_{RR}(p)\rangle &=& 
\sum_{n=0}^{\infty}
\langle G^{(2n)}_{RR}(p)\rangle = 
\sum_{n=0}^{\infty}
\frac{\delta^{2n}}{(\omega-pv)^{n+1}(\omega+pv)^n}
\label{MF sum}\\
&=& \frac{\omega+pv}{\omega^2-p^2v^2-\delta^2}.
\eeqa
The spectral function contains two narrow peaks,
\beqa
{\cal A}(\omega,{\bf p}) &=& -\pi^{-1}\mbox{Im}G_{RR}(\omega+i0,p)
\nonumber\\
&=& \frac{\tilde{\epsilon}_p+pv}{2\tilde{\epsilon}_p}\ 
\delta(\omega-\tilde{\epsilon}_p)
+ \frac{\tilde{\epsilon}_p-pv}{2\tilde{\epsilon}_p}\ 
\delta(\omega+\tilde{\epsilon}_p),
\label{BCS A}
\eeqa
where 
\beq
\tilde{\epsilon}_p =\sqrt{p^2v^2+\delta^2}
\eeq
(recall that $p$ is the distance to the Fermi momentum $\pm p_F$).  
The density of states vanishes for $|\omega|<\delta$ 
and exhibits a pile-up near $\omega=\pm\delta$:
\beq
{\cal N}(\omega) = \frac{\theta(\omega^2-\delta^2)}{\pi v}\ 
\frac{|\omega|}{\sqrt{\omega^2-\delta^2}}.
\label{BCS DOS}
\eeq

To describe a state without a long-range order, one can consider a phonon
statistics with a fixed gap amplitude $\delta$ and a fluctuating phase
(the nonlinear $O(2)\ \sigma$ model).  The model is characterized by a
single parameter, a temperature-independent phase stifness $\alpha$.
The two-point phonon correlation function has been calculated, e.g., in 
\cite{Gruner}:
\beq
D(x,x') = \delta^2 e^{-|x-x'|/\xi},
\hskip 1cm
\xi = \alpha/T.
\eeq
Higher-order correlations can be computed in a similar way: 
\beqa
&&D(x_n,x_n',\ldots,x_1,x_1') \equiv
\langle
\Delta(x_n)\,\Delta\!^*(x_n')\ldots
\Delta(x_1)\,\Delta\!^*(x_1')
\rangle
\nonumber\\
&& = \delta^{2n} 
\exp{\left(-\sum_{i,j=1}^{n}
\frac{|x_i-x_j'| + |x_i'-x_j| - |x_i-x_j| - |x_i'-x_j'|}{2\xi}\right)}.
\eeqa
As the temperature approaches zero, $\xi\to\infty$ and one recovers 
the statistics of the mean-field theory (\ref{MF statistics}).  Accordingly, 
the fermion energy spectrum in the limit of a long correlation length
approaches the BCS form (\ref{BCS DOS}), something one rightfully expects.  

\section{Model of Sadovskii}
\label{Sadovskii's model}

A phonon system with different statistics was considered in the
1970's by M.~V.~Sadovskii.  Instead of a fixed gap amplitude 
and Gaussian phase fluctuations (as in Sec.~\ref{beyond 2nd order}), his model
is concerned with independent Gaussian fluctuations of real and imaginary 
parts of $\Delta(x)$.  In other words, both phase {\em and amplitude}
of the gap are allowed to fluctuate.  This feature leads to a very 
different fermion spectrum in the limit of slow fluctuations.

\subsection{Phonons with Gaussian statistics}
\label{define exact}

Statistical properties of a Gaussian random variable $\Delta(x)$ are 
completely determined once the mean value and two-point correlations are 
specified:
\beqa
\langle\Delta(x)\rangle = 
\langle\Delta\!^*(x)\rangle &=& 0,
\label{mean}\\
\langle\Delta(x)\Delta(x')\rangle = 
\langle\Delta\!^*(x)\Delta\!^*(x')\rangle &=& 0,
\label{D anomalous}\\
\langle\Delta(x)\Delta\!^*(x')\rangle = D(x-x')
&\equiv& \delta^2 e^{-|x-x'|/\xi}. 
\label{D normal}
\eeqa
All higher-order correlations (\ref{D2n}) are then given by Wick's theorem,
\beqa
D(x_n,x_n',\ldots,x_1,x_1') &=& 
D(x_n-x_n') \ldots D(x_2-x_2')\,D(x_1-x_1') 
\nonumber\\
&& + \mbox{ permutations of primed coordinates.}
\label{n!}
\eeqa
The right-hand side includes $n!$ terms, e.g., 
\beq
D(x_2,x_2',x_1,x_1') = \delta^4 e^{-|x_2-x_2'|/\xi} e^{-|x_1-x_1'|/\xi}
+ \delta^4 e^{-|x_2-x_1'|/\xi} e^{-|x_1-x_2'|/\xi}.
\label{D4}
\eeq

\subsection{Solution for $\xi\to\infty$}

As first noted by Sadovskii \cite{Sad73}, 
the determination of the electron energy spectrum simplifies in the
limit of long-range (slow) fluctuations of the order parameter, 
$\xi\gg\delta/v$.  The electron energy spectrum in this limit is
strikingly different from the spectrum with a sharp gap discussed 
in Sec.~\ref{beyond 2nd order}.  Instead, one finds a broadly smeared gap,
or pseudogap, which is caused by fluctuations of the gap amplitude,
absent in the previous model.  

Taking $\xi\to\infty$ reduces (\ref{D normal}) to a
constant and thus (\ref{n!}) becomes coordinate-independent as well:
\beq
D(x_n,x_n',\ldots,x_1,x_1') = n!\ \delta^{2n}.
\eeq
By comparing this result to what we had for the state with a fixed
$\Delta$ (\ref{MF statistics}), we can write the following expression 
for the fermion propagator 
\beq
\langle G_{RR}(p)\rangle =
\sum_{n=0}^{\infty}
\langle G^{(2n)}_{RR}(p)\rangle = 
\sum_{n=0}^{\infty}
n!\ \frac{\delta^{2n}}{(\omega-pv)^{n+1}(\omega+pv)^n}.
\eeq
The only difference from (\ref{MF sum}) is the factor $n!$, which makes
the sum divergent for any frequency and momentum.  

This difficulty can be circumvented \cite{Sad73} if we recognize that
the divergent sum is an asymptotic expansion of the Stiltjes integral: 
\beq
\int_0^\infty\!\frac{e^{-t}dt}{1-tx} = \sum_{n=0}^\infty n!\, x^n.
\label{Stiltjes}
\eeq
This is precisely our series.  The left-hand side is perfectly 
finite for any $x$ away from the positive real axis.  If $x$ approaches
the real axis from the complex plane, $x\pm i0$, the integral (\ref{Stiltjes})
has a non-zero imaginary part, not reproducible by a sum of positive
numbers on the right-hand side, hence a divergence. 

Rather than trying to resum a divergent series, it is more useful to 
remove the divergence all together.  For that purpose, we will go back 
to the original assumption about the Gaussian statistics, which is the 
source of $n!$.  In the limit $\xi\to\infty$, 
instead of a random field $\Delta(x)$, we have a single random variable 
$\Delta$ describing the value of the gap field everywhere on the chain.  Its 
Gaussian character, postulated above, is realized by considering an ensemble
of chains, each with a different but {\em fixed} $\Delta$, with the 
distribution (``density of chains'')
\beq
\rho(\Delta) = e^{-|\Delta|^2/\delta^2}/\pi\delta^2,
\label{gap density}
\eeq
which gives, as required, 
\beq
\langle\Delta^n{\Delta\!^*}^n\rangle \equiv 
\int\!|\Delta|^{2n}\ \rho(\Delta)\, d^2\Delta
= n!\,\delta^{2n}.
\eeq
On every single chain, there is a perfect Peierls gap of size $|\Delta|$,
which, however, varies from chain to chain.  

To obtain, e.g., the density 
of states in such an ensemble, one can average the result for a 
single gap (\ref{BCS DOS}) over the distribution of gaps (\ref{gap density}),
which gives a smeared-out gap \cite{Sad73}.  The density of states vanishes as 
$\omega^2$ at low frequencies.  
The fermion spectral function can be obtained in a similar way, by 
integrating the BCS spectral weight with two $\delta$-functions (\ref{BCS A}) 
over the gap distribution (\ref{gap density}).  As a result of a varying 
gap amplitude 
$|\Delta|$, one finds \cite{Sad73} peaks that are significantly broad,
especially near the Fermi points ($p=0$), where the spectral line shape is
\beq
{\cal A}(\omega,0) = |\omega|\delta^{-2}e^{-\omega^2/\delta^2}.
\eeq
A large linewidth, of order $\delta$, reflects not a scattering rate, 
but rather an inhomogeneous broadening due to a varying gap size.  

\subsection{What do Sadovskii's chains look like?}

\begin{figure}[t!b!]
\begin{center}
\leavevmode
\epsfxsize 4in
\epsffile{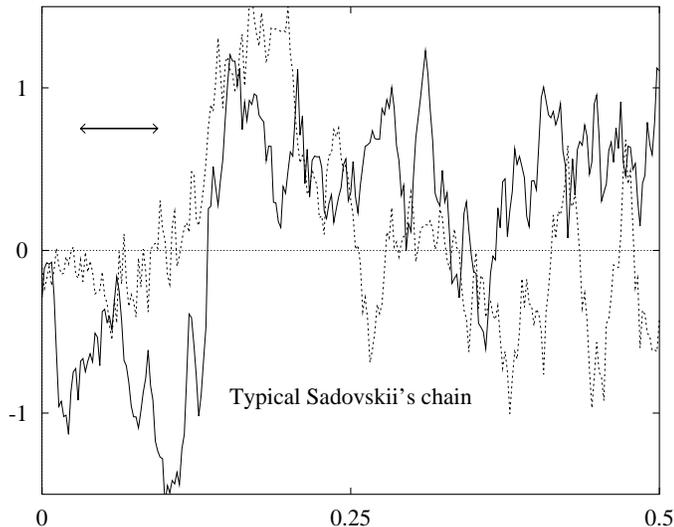}
\end{center}
\caption{Half of a typical Sadovskii chain.  The correlation length 
(the two-head arrow) is $\xi=1/8$, the mean value of the gap is 
$\delta=1$.  Solid line: $\mbox{Re}\Delta(x)$, 
dashed line: $\mbox{Im}\Delta(x)$.  The actual atomic displacement
at a point $x$ is given by $\mbox{\rm Re}\left[\Delta(x)e^{2ik_Fx}\right]$.}  
\label{Sad chain}
\end{figure}

A typical Sadovskii's chain is shown in Fig.~\ref{Sad chain}.  The gap
field, or the complex amplitude of atomic displacements, 
remains approximately constant over distances smaller than the correlation 
length $\xi$.  Even though this chain looks very rough, its two-point 
correlation function $D(x,x')$ is quite smooth.  
In fact, 
\beq
D(x,x') = \delta^2 e^{-|x-x'|/\xi}
\label{ideal D2}
\eeq
{\em exactly} for this particular chain.  

But wait.  How is a correlation function defined for a {\em single} chain?
And why is the correlation function translation invariant, 
\beq
D(x+\zeta,x'+\zeta) = D(x,x'),
\eeq
whereas the chain is not?  Answer:
\beq
D(x,x')\ \stackrel{{\rm def}}{=}\ 
\frac{1}{L}\int_0^L\!d\zeta\ \Delta(x+\zeta)\, \Delta\!^*(x'+\zeta).
\label{D2 defined}
\eeq
This is obviously translation invariant for periodic boundary conditions.  

To see how a chain with the right correlation function can be constructed,
rewrite (\ref{D2 defined}) in terms of Fourier components 
\beqa
D(x,x') &=& L^{-1}\int_0^L\!d\zeta \ \ 
L^{-1/2}\sum_{k} \Delta_k e^{ik(x+\zeta)} \ \ 
L^{-1/2}\sum_{k'}\Delta\!^*\,_{k'} e^{-ik'(x'+\zeta)}
\nonumber\\
&=& L^{-1}\sum_k |\Delta_k|^2 e^{ik(x-x')}
\to \int\!\frac{dk}{2\pi}\ |\Delta_k|^2 e^{ik(x-x')}
\label{D2 Fourier}
\eeqa
in the limit $L\to\infty$.  Choosing now 
\beq
\Delta_k = \delta e^{i\theta_k} 
\sqrt{\frac{2\xi^{-1}}{k^2+\xi^{-2}}}
\label{Delta_k}
\eeq
with an arbitrary phase $\theta_k$ immediately yields (\ref{ideal D2}).  
This is approximately how the chain in Fig.~\ref{Sad chain}
has been simulated.  

Furthermore, it can now be seen that an {\em ensemble} of such chains 
exhibits the required Gaussian statistics (\ref{n!}).  For instance, the 
four-point correlation function for a {\em single} chain is
\beqa
D(x_2,x_2',x_1,x_1') 
&\stackrel{{\rm def}}{=}&
\frac{1}{L}\int_0^L\!d\zeta\ 
\Delta(x_2+\zeta)\, 
\Delta\!^*(x_2'+\zeta)
\Delta(x_1+\zeta)\, 
\Delta\!^*(x_1'+\zeta)
\nonumber\\
&=& L^{-2}\sum_{\{k\}}
\Delta_{k_2}e^{ik_2 x_2}\,
\Delta\!^*\,_{k_2'}e^{-ik_2' x_2'}\,
\Delta_{k_1}e^{ik_1 x_1}\,
\Delta\!^*\,_{k_1'}e^{-ik_1' x_1'},
\label{D4 defined}
\eeqa
where momenta satisfy the constraint $k_1+k_2=k_1'+k_2'$. 
 
The crucial step is to average (\ref{D4 defined}) over the arbitrary phases
$\theta_k$, which enter this expression in the form of the factor
\beq
\exp{\left[i(\theta_{k_1}+\theta_{k_2}-\theta_{k_1'}-\theta_{k_2'})\right]}.
\label{pre-Gaussian}
\eeq
Phases $\theta_k$ are independent random variables uniformly distributed 
in the interval $0<\theta_k<2\pi$.  Averaging over them makes 
(\ref{pre-Gaussian}) vanish, unless the phase factors cancel one another 
pairwise:
\beq
{k_1}={k_1'}, \ \ 
{k_2}={k_2'}, 
\hskip 1cm \mbox{or} \hskip 1cm 
{k_1}={k_2'}, \ \ 
{k_2}={k_1'}. 
\label{survivors}
\eeq
The so averaged four-point correlation function reads
\beqa
\langle D(x_2,x_2',x_1,x_1') \rangle &=& 
L^{-1}\sum_{k_2}|\Delta_{k_2}|^2 e^{ik_2(x_2-x_2')}\ \ 
L^{-1}\sum_{k_1}|\Delta_{k_1}|^2 e^{ik_1(x_1-x_1')}
\nonumber\\
&+&
L^{-1}\sum_{k_2}|\Delta_{k_2}|^2 e^{ik_2(x_2-x_1')}\ \ 
L^{-1}\sum_{k_1}|\Delta_{k_1}|^2 e^{ik_1(x_1-x_2')}
\label{almost Gaussian}\\
&-&
L^{-2}\sum_{k}|\Delta|^4 e^{ik(x_2-x_2'+x_1-x_1')}.
\nonumber
\eeqa
The third line in (\ref{almost Gaussian}) is needed to adjust for
the overcounting of the terms with $k_1=k_1'=k_2=k_2'$ in (\ref{survivors}).
Comparing these to (\ref{D2 Fourier}) reveals
(almost) Gaussian statistics:
\beqa
\langle D(x_2,x_2',x_1,x_1') \rangle &=& 
D(x_2,x_2')\, D(x_1,x_1') + 
D(x_2,x_1')\, D(x_1,x_2') 
\\
&-&
L^{-1}\int\!\frac{dk}{2\pi}|\Delta_k|^4 e^{ik(x_2-x_2'+x_1-x_1')}.
\nonumber
\eeqa
In the limit $L\to\infty$, the extra term vanishes as $\xi/L$ or faster.  
This procedure can evidently be extended to arbitrarily high orders.

\begin{figure}[tb]
\begin{center}
\leavevmode
\epsfxsize 4in
\epsffile{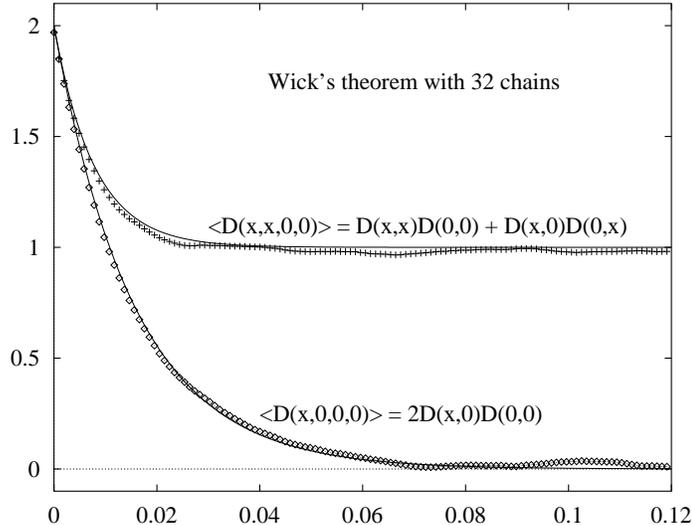}
\end{center}
\caption{Four-point correlation functions $\langle D(x,x,0,0) \rangle$
(crosses) and $\langle D(x,0,x,0)\rangle$ (diamonds) averaged over 32
different chains.  Solid lines are the corresponding Gaussian curves.  
$\xi=L/64$, $\delta=1$. }
\label{Gaussian nature}
\end{figure}

We thus have found an efficient way to simulate the ensemble considered
by Sadovskii.  Namely, by using the known Fourier amplitudes 
(\ref{Delta_k}), we can generate a sufficiently large number of chains with 
different choices of random phases $\theta_k$.  That this number does not
have to be too large is demonstrated in Fig.~\ref{Gaussian nature}.  
Average four-point correlations in an ensemble of just 32 chains
agree quite well with their Gaussian expectation values.  
Note a systematic downward shift for $\langle D(x,x,0,0) \rangle$,
which is expected to equal $\xi/L$.  

As a matter of fact, 
it is not necessary to consider an ensemble of chains.  Instead, one 
can regard a single long enough chain as an ensemble of its segments
(of length $L$).  Doing so resolves the apparent paradox of 
strongly broadened electron states in the limit $\xi\to\infty$.  
As long as one studies properties of electrons within a 
{\em single} segment, this broadening will exist until $\xi$ exceeds
the segment length $L$.  In the limit $\xi\gg L$, electrons will 
see a well-defined gap $\Delta$ (within this segment) and their 
spectral function will exhibit two sharp Bogoliubov peaks 
at energies $\pm\sqrt{p^2v^2+|\Delta|^2}$.  However, if we now pack our
instruments and go to another segment of the chain, far enough away,
we could find there a different 
value of the gap.  Averaging over an {\em infinitely long} chain,
on the other hand, will always give the Gaussian broadening as 
$\xi < L = \infty$ always in that case.  

{\em Remark. }
The possibility of gap-amplitude fluctuations is largely ignored in
the literature.  Common-sense wisdom suggests that variations 
of the energy gap cost too much energy and therefore only the phase 
of the order parameter $\Delta$ is expected to fluctuate at low temperatures.
There are two counterarguments here.  First, a qualitative one, that 
gap-size fluctuations further increase the entropy and thus reduce the 
{\em free energy}.  Another, quantitative argument appeals to the well-known 
solution \cite{Luther} for the fermion spectrum in the Luttinger model with 
attraction, truly an exactly solvable model with a pseudogap.  
Cooper pairs, the dominating fluctuations in this model \cite{Luther}, 
produce a broadly smeared energy gap in the fermion spectrum, even at zero 
temperature.  In the limit of a large correlation length, phase 
fluctuations alone cannot account for a strong smearing.  Thus, 
amplitude fluctuations are inevitably present in this model and may
be more commonplace than usually thought.

\section{Sadovskii's solution}
\label{Sadovskii wrong}

We now return to the theoretical analysis of the problem, this time for 
a finite correlation length $\xi$.  

\subsection{Sadovskii's conjecture}

As previously noted, evaluation of 
Feynman diagrams is more convenient in momentum space, where
the two-point phonon correlation is a Lorentzian (\ref{Lorentzian line}),
\beq
D(k) = \delta^2\frac{2\xi^{-1}}{k^2+\xi^{-2}}
\eeq
As we have seen before, the order-$\delta^2$ correction to the electron
Green's function is (Fig.~\ref{LRA})
\beq
G_{RR}^{(2)}(p) 
= \delta^2 \frac{1}{\omega-p+i0}\ 
\frac{1}{\omega+p+i\xi^{-1}}\ 
\frac{1}{\omega-p+i0}.
\label{conjecture}
\eeq
This line is rather transparent: integration over momentum $k$ transfered
to the phonon simply shifts the imaginary in the denominator of the 
intermediate electron propagator from $+i0$ to $+i\xi^{-1}$.  

\begin{figure}
\begin{center}
\leavevmode
\epsfxsize 5in
\epsffile{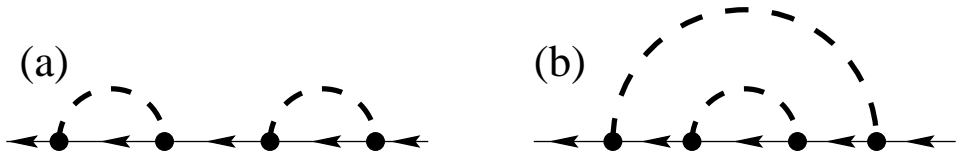}
\end{center}
\caption{Two-phonon contributions to the fermion propagator. 
(a) is generated by the first-order self-energy, while (b) 
contains two-phonon self-energy.}
\label{second order}
\end{figure}

Such a simple form of the second-order
correction,
\beq
\Sigma^{(2)}_{RR}(\omega,p)\propto G^{(0)}_{LL}(\omega+i\xi^{-1},p), 
\eeq
has prompted Sadovskii to 
{\em conjecture} that contributions of higher-order graphs 
to $G_{\sigma\sigma}(p)$ are given by the following simple rules: 
\beqa
&&\mbox{A phonon line contributes $\delta^2$}, \\
&&\mbox{An electron line contributes }
\frac{1}{\omega \pm p + i\nu \xi^{-1}},
\label{Ansatz}
\eeqa
where $\nu$ is the number of phonon lines above a given electron line.  
The sign in front of $p$ alternates as the fermion propagates left and right.  

For instance, according to this rule, corrections of order $\delta^4$ 
to the electron Green's function (Fig.~\ref{second order}) should read
\beqa
\mbox{(a)} &=& \delta^4 \left(\frac{1}{\omega-p+i0}\right)^3
\left(\frac{1}{\omega+p+i\xi^{-1}}\right)^2,
\label{4(a)}\\
\mbox{(b)} &=& \delta^4 \left(\frac{1}{\omega-p+i0}\right)^2
\frac{1}{\omega-p+2i\xi^{-1}}\ 
\left(\frac{1}{\omega+p+i\xi^{-1}}\right)^2.
\label{4(b)}
\eeqa 

Basing on this {\em Ansatz}, Sadovskii was able to derive and solve a 
recursion relation for the self-energy of order $\delta^{2n}$ 
\cite{Sad79} following a method due to Elyutin \cite{Elyutin}.  
The exact Green's function was then obtained in a
continued fraction representation.  This remarkable derivation is 
getting quite popular these days \cite{McKenzie,Schmalian}.  

\subsection{Failure in order $\delta^4$}

Unfortunately, {\em Ansatz} (\ref{Ansatz}) works only for a limited 
class of diagrams [e.g., Fig.~\ref{second order}(a)] and is simply
{\em incorrect} for others [Fig.~\ref{second order}(b)].  The problem, 
quite mundane, is in sloppy handling of the imaginary
part  --- equal to $+i0\,\mbox{sign}(\omega)$ or $i\omega_n$ depending 
on the formalism --- in the denominator of $G^{(0)}(\omega,p)$.  

Recall that, to order $\delta^2$, we integrated 
\beq
\int\!\frac{dk}{2\pi}
\frac{1}{\omega+p-k+i0}\ 
\frac{2\xi^{-1}}{k^2+\xi^{-2}},
\eeq
which has two poles above the real $k$ axis and only one pole below.  
If we complete the integration contour in the lower half of the plane, 
only one pole is inside and the resulting expression is simple.  

In evaluating graph (b) in Fig.~\ref{second order}, the integral over
momentum $q$ of the external phonon line reads
\beq
\int\!\frac{dq}{2\pi}\ 
\frac{1}{[\omega+(p-q)+i0]^2}\ 
\frac{1}{\omega-(p-q-k)+i0}\ 
\frac{2\xi^{-1}}{q^2+\xi^{-2}}, 
\eeq
which has two poles on either side of the real $q$ axis, so that, whichever way
the contour is completed at infinity, the result contains two terms, 
rather than one.  Integrating over $k$ first does not help either:
\beq
\int\!\frac{dq}{2\pi}\ 
\frac{1}{[\omega+(p-q)+i0]^2}\ 
\frac{1}{\omega-(p-q)+i\xi^{-1}}\ 
\frac{2\xi^{-1}}{q^2+\xi^{-2}}
\label{after k}
\eeq
is plagued by the same problem.  The integral over $q$, completed above the
real axis, yields the result conjectured by Sadovskii (\ref{4(b)}) 
{\em plus} a non-zero contribution from the pole at $q=\omega+p+i0$.  

The situation does not change when one uses thermal Green's functions,
in which case $\omega+i0\,\mbox{sign}(\omega)$ is replaced with $i\omega_n$ 
and the same problem arises.  In higher orders, expressions for 
$G_{\sigma\sigma}^{(2n)}(p)$ becomes progressively more complicated
by the presence of diagrams with a phonon line running over many electron 
propagators.  

\subsection{Which problem did Sadovskii solve, exactly?}
\label{Which model}

\begin{figure}
\begin{center}
\leavevmode
\epsfxsize 5.5in
\epsffile{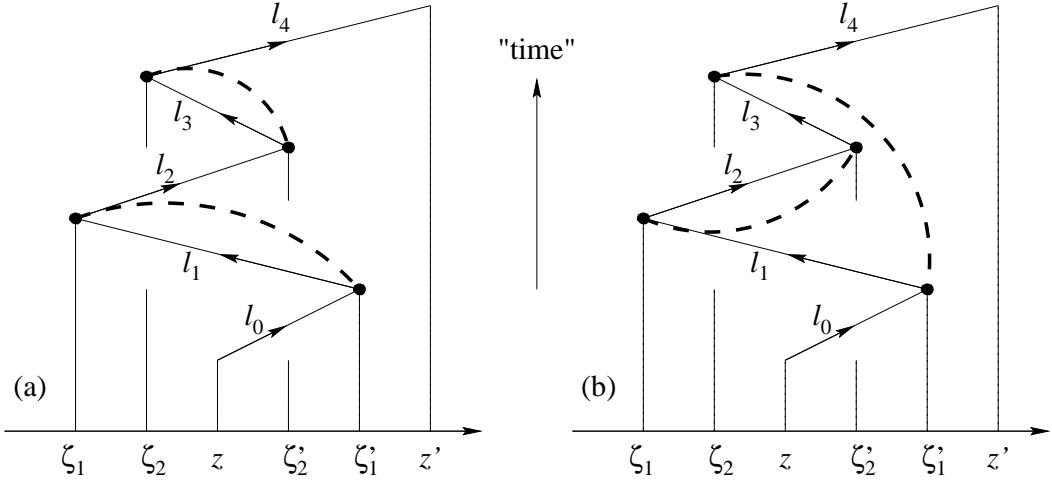}
\end{center}
\caption{Correction to the fermion propagator $G^{(4)}(x',x)$.
Solid lines: free fermion propagator $-ie^{i\omega l_n}$.  
Dashed lines: two-point phonon correlation 
$\delta^2e^{-|\zeta_i-\zeta_j'|/\xi}$.  Vertical dimension is added
for clarity.  
}
\label{paths4}
\end{figure}

The fact that the trouble is caused by infinitesimal imaginary numbers
in fermion propagators may create an illusion that the problem can
be somehow fixed.  It is more instructive to look at it in 
configuration space.  We will now see exactly which problem Sadovskii solved.  

Using conventions of Sec.~\ref{G to 2nd order}, we write out the expression 
for the first of the two diagrams for $\langle G_{RR}^{(4)}(p)\rangle$, 
Fig.~\ref{paths4}(a):
\beqa
&&-i\int_{0}^\infty \!\!dl_4\ldots dl_0\, 
e^{i(\omega-p)l_4}
e^{i(\omega+p)l_3}
e^{i(\omega-p)l_2}
e^{i(\omega+p)l_1}
e^{i(\omega-p)l_0}
\nonumber\\
&&\times\ \delta^4\ e^{-|\zeta_2-\zeta_2'|/\xi}\ e^{-|\zeta_1-\zeta_1'|/\xi}
\label{expression 4(a)}
\eeqa
[cf.~Eq.(\ref{G2(x,x')})].  As $|\zeta_2-\zeta_2'|=l_3$ and 
$|\zeta_1-\zeta_1'|=l_1$, the integrals over lengths $\{l_n\}$ can
be immediately carried out and one obtains (\ref{4(a)}).  

The other diagram, Fig.~\ref{paths4}(b), differs by a permutation of
$\zeta_1'$ and $\zeta_2'$, so that only the second line of 
(\ref{expression 4(a)}) changes and now reads
\beq
\times\ \delta^4\ e^{-|\zeta_2-\zeta_1'|/\xi}\ e^{-|\zeta_1-\zeta_2'|/\xi}.
\label{expression 4(b)}
\eeq
While $|\zeta_1-\zeta_2'|=l_2$, the other distance, $|\zeta_2-\zeta_1'|$,
cannot be simply expressed as a sum of some path lengths, which is
what causes the problem.  Note, however, that, had we replaced the 
physical distance $|\zeta_2-\zeta_1'|$ with the sum of path lengths 
$l_1+l_2+l_3$, the previous expression would have read
\beq
\times\ \delta^4\ e^{-l_1/\xi}\ e^{-2l_2/\xi}\ e^{-l_3/\xi},
\label{expression 4(b) a la Sad}
\eeq
which could be easily integrated over lengths yielding Eq.~(\ref{4(b)}),
precisely what Sadovskii wanted.  

Once the physical distance between two points $|\zeta_i-\zeta_j'|$ 
in the phonon correlation function $D(\zeta_i,\zeta_j')$
has been replaced with the length of the fermion path between these points, 
Sadovskii's conjecture (\ref{conjecture}) is valid in all orders of 
perturbation theory.  Indeed, define $\nu_m$ to be the number of phonon 
lines above the fermion leg $l_m$, which can be done unamiguously by 
straightening out the fermion trajectory (i.e., by using 
Fig.~\ref{second order} instead of Fig.~\ref{paths4}).  
The contribution of a given diagram to 
$G^{2n}_{RR}(p)$ will then be a product of independent factors 
\beqa
\delta^{2n}\prod_{m=0}^{2n} (-i) \int_0^{\infty}\!
e^{i\omega l_m} e^{-i(-1)^m p l_m} e^{-\nu_m l_m/\xi} dl_m
\nonumber \\
= \delta^{2n}\prod_{m=0}^{2n} \frac{1}{\omega - (-1)^m p +i\nu_m\xi^{-1}},
\eeqa
precisely as required by (\ref{conjecture}).  

It is thus clear that the original {\em Ansatz} of Sadovskii solves 
a rather unphysical problem, in which phonon correlations 
$\langle\Delta(x)\Delta\!^*(x')\rangle$ depend not on the geometrical
distance $|x-x'|$, but rather on the length of the path the fermion 
traveled between points $x$ and $x'$.  This point is further 
illustrated using a {\em two}-dimensional example in Sec.~\ref{altius}.  

\section{Extension to higher dimensions?}
\label{altius}

It has already been mentioned that the calculation of Sadovskii 
is tailored to one spatial dimension.  This limitation stems from 
the fact that the order-$\delta^2$ correction to the fermion self-energy 
in the presence of classical fluctuations of an order parameter,
\beq
\int\frac{d^d{\bf k}}{(2\pi)^d}\ 
\frac{1}{\omega+{\bf(p-k)\!\cdot\!v}+i0}\ \frac{2\xi^{-1}}{k^2+\xi^{-2}}, 
\label{Sigma in d dim}
\eeq
has a simple form in $d=1$ dimension only.  Not having a simple,
``self-replicating'' form for the lowest-order correction possibly 
indicates that there is little hope of finding a general recipe for 
higher orders.

\subsection{Model of ``hot spots'' in the cuprates}
\label{hot spots}

Recently, however, J.~Schmalian, D.~Pines and B.~Stojkovi\'c \cite{Schmalian} 
applied the ideas of Sadovskii to a {\em two}-dimensional system, high-$T_P$ 
cuprate superconductors, to investigate the nearly antiferromagnetic
Fermi liquid (NAFL) \cite{NAFL}.  
This development further illustrates 
in what context the solution of Sadovskii is applicable.  It turns out
that the dimensionality of the system is not important.  
A really necessary ingredient is the peculiar form of order-parameter 
correlations, which should decay exponentially with the ``distance''
measured {\em along the fermion path}.  
(There is also a technical, but very important, requirement that the 
spectrum of free electrons be flat, i.e., 
$\epsilon_{{\bf p}_F+{\bf p}} = {\bf p\!\cdot\!v}$, where $\bf v$ is 
a constant vector.)  

\begin{figure}[t]
\begin{center}
\leavevmode
\epsfxsize 3.5in
\epsffile{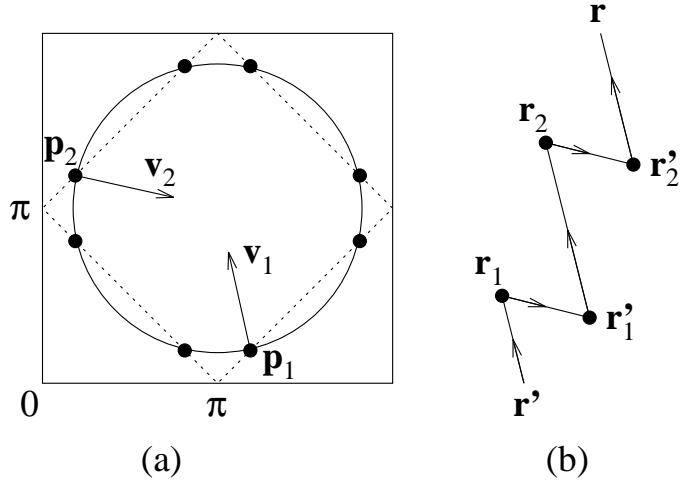}
\end{center}
\caption{(a)  A sketch of the Fermi surface (solid line) in the cuprates.  
${\bf v}_1$ and ${\bf v}_2$ are Fermi velocities at two hot spots 
(filled circles) connected by the antiferromagnetic wave vector 
${\bf Q}=(\pi,\pi)$.  The dashed line is the locus of states most 
strongly affected by the AFM scattering, 
$\epsilon_{\bf p+Q}=\epsilon_{\bf p}$.  (b) A fermion 
initially in the vicinity of 
the hot spot ${\bf p}_1$ travels in a zigzag manner in real space switching 
between non-collinear velocities ${\bf v}_1$ and ${\bf v}_2$ as it is 
scattered by spin fluctuations.  
}
\label{Schmal path}
\end{figure}

In the NAFL approach, electrons, considered to be ideal fermions, interact 
with antiferromagnetic (AFM) spin fluctuations, whose static susceptibility
is peaked near wavenumber ${\bf Q}=(\pi,\pi)$ in reciprocal lattice units:
\beq
\chi({\bf Q+q}) \approx \frac{\chi({\bf Q})}{1+q^2\xi^2}.
\label{NAFL chi}
\eeq
The strongest effect of AFM fluctuations on the fermion energy spectrum is 
expected when
scattering by wave vector ${\bf Q}$ connects states of the same energy,
$\epsilon_{\bf p+Q}=\epsilon_{\bf p}$.  Such points in the Brillouin zone
form a line shown in Fig.~\ref{Schmal path}(a) for a tight-binding fermion 
energy spectrum (nearest and next-nearest neighbor hopping).  Places where
this line intersects the Fermi surface have been termed ``hot spots''.  
Low-energy fermionic excitations in these spots are
presumably fried by spin fluctuations and are short-lived, hence the name.  
This must be true, at least {\em to some extent}, as photoemission shows 
extremely broad peaks (hundreds of meV) in the electron spectral weight 
${\cal A}(\omega,{\bf p})$ at these momenta \cite{broad ARPES}.  

Consider the lowest-order fermion self-energy from one-magnon exchange 
(the same diagram as in Fig.~\ref{LRA}).  
After linearizing the free fermion spectrum near the hot spots
\beq
\epsilon_{{\bf p}_n+{\bf p}}\approx {\bf v}_n\!\cdot\!{\bf p},
\label{linearization}
\eeq
the self-energy for a fermion near hot spot ${\bf p}_1$ reads
\beq
\Sigma(\omega,{\bf p}_1+{\bf p}) \approx 
\int\frac{d^2 {\bf q}}{(2\pi)^2}\ 
\frac{\chi({\bf Q})}{1+q^2\xi^{-2}}\ 
\frac{1}{\omega + {\bf v}_2\!\cdot\!({\bf p-q})+i0}.
\eeq
Here ${\bf Q+q}$ is the momentum transfered to the magnon.  Note that the 
intermediate electron is near the other hot spot ${\bf p}_2$.  This 
has pecisely the form of Eq.~(\ref{Sigma in d dim}) and one cannot 
get a simple expression out of it, to say nothing of higher-order corrections.

Schmalian et al.~noted that Fermi velocities at conjugated hot spots 
(e.g., ${\bf v}_1$ and ${\bf v}_2$) are almost perpendicular to each other.
If then one replaces the susceptibility (\ref{NAFL chi}) with a product 
\beq
\chi({\bf Q+q}) \approx \delta^2
\frac{2\xi^{-1}}{q_1^2+\xi^{-2}}\ 
\frac{2\xi^{-1}}{q_2^2+\xi^{-2}}, 
\label{chi factorized}
\eeq
where $q_n$ is the component of ${\bf q}$ along ${\bf v}_n$, a very
simple sef-energy results:
\beq
\Sigma(\omega,{\bf p}_1+{\bf p}) \approx 
\frac{\delta^2}{\omega+ {\bf v}_2\!\cdot\!{\bf p}+iv\xi^{-1}}
= G^{(0)}(\omega+iv\xi^{-1}, {\bf p}_2 + {\bf p}),
\eeq
where $v=|{\bf v}_1|=|{\bf v}_2|$.  Moreover, higher-order diagrams 
can be evaluated in a similar manner yielding simple expressions 
in the form conjectured by Sadovskii (\ref{conjecture}).  Lo and behold,
the problem becomes tractable to arbitrary order and the electron
Green's function can be obtained in the continued fraction 
representation \cite{Schmalian}, as discussed by Sadovskii.  

\subsection{What makes it solvable}

One should not be surprised that the trick with factorization 
(\ref{chi factorized}) makes the problem solvable.  The factorization
amounts to taking spin-spin correlations in real space in the form
\beq
\chi({\bf r-r'}) \propto \langle s^+({\bf r}) s^-({\bf r'})\rangle 
\propto e^{-|x-x'|/\xi} e^{-|y-y'|/\xi},
\eeq
where $x$ and $y$ are Cartesian components of the electron in the plane
(along the directions of ${\bf v}_1$ and ${\bf v}_2$, i.e., approximately
along the crystal axes).  This is precisely the same as to say that 
order parameter correlations decay with the ``distance'' measured 
along the fermion path, Fig.~\ref{Schmal path}(b), as I noted in the 
beginning of this Section.  That said, it is not even necessary to 
require that ${\bf v}_1$ and ${\bf v}_2$ be orthogonal.  

Technically, the similarity with the one-dimensional problem of Sadovskii 
arises because the electron energy spectrum has been linearized, 
whatever the actual number of dimensions is.
Indeed, according to (\ref{linearization}), the energy as a function 
of momentum varies only in the direction of ${\bf v}_n$.  The fermion
spectrum is {\em exactly} dispersionless in all directions 
perpendicular to ${\bf v}_n$.  This means, literally, that a free fermion 
propagates along a straight line, Fig.~\ref{Schmal path}(b).
In this approximation, $d^2\epsilon/dp_i dp_j=0$, a localized 
wave packet does not disperse as it propagates with velocity 
$v_i = d\epsilon/dp_i$.  It would not be a stretch to say that 
this is essentially a one-dimensional problem.  

\subsection{No pseudogap in the DOS}

Despite great similarities, there is one important aspect in which 
this two-dimensional
problem differs from the purely one-dimensional case of Sadovskii.  
In plain English, a fermion {\em never} returns to a starting point:
it zigzags away, Fig.~\ref{Schmal path}(b).  In contrast, 
${\bf v}_1=-{\bf v}_2$ in one dimension and a fermion {\em does} return to
the starting point ``once in a while''.  An important consequence of
this innocuous observation is that, with ${\bf v}_1\neq-{\bf v}_2$, 
the local fermion propagator is unaffected by fluctuations to all 
orders,
\beq
G(\omega,x,x) = G^{(0)}(\omega,x,x),
\eeq
because $G^{(n)}(\omega,x,x)=0$ for any order $n>0$.  Therefore, the 
local density of electron states is exactly the same as for free, 
noninteracting fermions:
\beq
{\cal N}(\omega) 
= -\pi^{-1}\mbox{Im}G(\omega,x,x) 
= -\pi^{-1}\mbox{Im}G^{(0)}(\omega,x,x) 
= {\cal N}^{(0)}(\omega). 
\eeq
In particular, this means that a {\em local} probe, such as tunneling
microscopy \cite{Renner} or NMR \cite{NMR}, should {\em not} observe any 
pseudogap behavior in the ``hot spots'' scenario!  

The use of the coordinate representation makes a proof of this statement
almost trivial: the step function in the free electron propagator 
(\ref{G0(omega,x)}) makes all corrections to the free propagator
vanish for fermion trajectories with points outside the cone formed by the 
vectors ${\bf v}_1$ and ${\bf v}_2$.  Thereby trajectories returning to the 
starting point consist of a single point, have zero integration 
measure (in the case of at least one intermediate point) and therefore do not 
contribute to the propagator.  

Of course, this can be seen in momentum space as well.  For simplicity,
take the velocities ${\bf v}_1$ and ${\bf v}_2$ to be orthogonal to each 
other and choose a pair of 
coordinate axes along them.  A generic correction of order $\delta^{2n}$ 
to the propagator of a fermion near Hot Spot 1 contains two 
momentum-dependent factors: 
\beq
\delta^{2n} \prod_{a=0}^{n}\frac{1}{\omega-p_1+i\nu_{1a}\xi^{-1}}\ 
\prod_{b=1}^{n}\frac{1}{\omega-p_2+i\nu_{2b}\xi^{-1}},
\eeq
where $\nu_{1a},\nu_{2b}>0$ (strict inequality!).  Its contribution to the 
local DOS is obtained by integrating over the momentum components 
$p_1$ and $p_2$ and taking the imaginary part.  This expression
is an analytical function of $p_1$ below the real axis.  
If $n>0$, the integration contour can be completed at infinity in the lower 
half of the complex $p_1$ plane (the integrand vanishes there fast enough).
Since no singularities are encircled, the integral 
vanishes for any $n>0$.  The exceptional case $n=0$ (free propagator) has 
been dealt with in Sec.~\ref{G to 2nd order}.  

It is worth stressing that Sadovskii's solution should be considered as 
a long-wavelength approximation only (as it is based on a linearized electron 
spectrum).  In practical terms, one should {\em not}
attempt to draw conclusions about the detailed band structure basing on
a solution of this type.  Things like a Brillouin zone or 
a van Hove singularity simply do not belong in this theory.  While integrating
the spectral weight over a Brillouin zone may show a slight reduction 
in the density of states near the Fermi level \cite{slight_reduction}, 
such an extrapolation of an effective field theory to real-life details 
is not warranted.  The only conclusion that can be drawn safely is that,
as the lattice spacing is taken to zero, any trace of the pseudogap 
disappears.  Thus, the pseudogap is not natural in this model.  

\begin{figure}[t]
\begin{center}
\leavevmode
\epsfxsize 4in
\epsffile{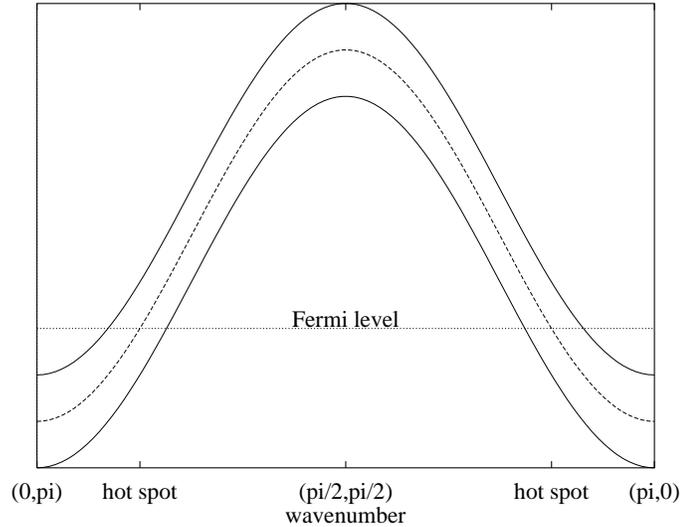}
\end{center}
\caption{Splitting of the free electron band (dashed line) into 
upper and lower bands (solid lines) in the presence of long-range 
AFM order with wavevector ${\bf Q} = (\pi/a,\pi/a)$.  When the band
splitting $2\delta$ is smaller than the bandwidth, there is no gap 
in the density of states.}
\label{nogap}
\end{figure}

While it may appear paradoxic that the DOS is unaffected, it is, in fact, 
a direct consequence of the assumptions that made the calculation of 
Schmalian et al.~possible.  It is also directly related 
to an observation by Randeria \cite{not tied} that the NAFL pseudogap 
is not tied to the Fermi surface (the dotted line and the solid line in 
Fig.~\ref{Schmal path}(a), respectively).  In the antiferromagnetic 
scenario for the pseudogap, the spectral weight of the fermion states
on the dotted line is moved from $\epsilon_{\bf p}$ to higher and lower 
energies in the range $\epsilon_{\bf p}\pm\delta$.  Since, however, 
the energy $\epsilon_{\bf p}$ varies along the dotted line (by the amount
equal to $4t_2$, where $t_2$ is the next-nearest neighbor hopping amplitude), 
the pseudogap will be completely washed out if $4t_2$ exceeds $\delta$.
Linearization of spectrum (\ref{linearization}) is equivalent to 
assuming $\delta\ll 4t_2$ (no local pseudogap).  
As Monthoux and Pines suggested, $4t_2=0.45$ eV \cite{Monthoux}, 
so that any pseudogap of a lesser width will be washed out in 
the density of states.  This situation is illustrated schematically 
in Fig.~\ref{nogap}.  

In contrast, there is no washing out of a pseudogap created 
by Cooper pair fluctuations.  In that case, fermion states 
coupled by emission or absorption of a Cooper pair are 
electrons and holes of equal momenta and spin.  Therefore, their
velocities are equal and opposite, 
\beq
{\bf v}_h = \frac{d(-\epsilon_{\bf -p})}{d{\bf p}} 
= - \frac{d\epsilon_{\bf p}}{d{\bf p}} 
= - {\bf v}_e,
\eeq
as long as time reversal is a good symmetry of the system
($\epsilon_{\bf -p}=\epsilon_{\bf p}$).  In the problem with a 
linearized dispersion, a fermion moves along a straight line 
back and forth alternating between an electron and a hole. 
There are non-zero corrections to the local propagator 
$G(\omega,x,x)$ in all orders, which means that scattering by
Cooper pair fluctuations does affect the local DOS.  Put simply, 
a pseudogap created by pairing fluctuations {\em is} tied to a Fermi surface.

\section{Summary}

In this paper, it has been demonstrated that the issue of phonon 
statistics is quite important for the properties of electrons in the 
pseudogap regime above the ordering temperature.  Knowledge of
the two-point correlation function $\langle\Delta(x)\,\Delta\!^*(x')\rangle$ 
allows one to compute the electron Green's function or self-energy 
to the second order in the gap size $\delta$ only.  When the 
correlation length of the fluctuations increases beyond the point 
$\xi>v/\delta$, higher-order phonon contributions become important,
which is why multi-phonon correlation functions are needed.  
It has been shown explicitly that different choices of phonon 
statistics lead to widely different results for the fermion spectrum
in the particularly interesting limit of slow fluctuations, 
$\xi\gg v/\delta$.  

A model of phonons with Gaussian staistics \cite{Sad73} has been 
revisited and thoroughly discussed, both in momentum and coordinate 
domains.  It has been shown that its ``exact'' solution 
for a finite correlation length \cite{Sad79} contains an error and, 
in fact, solves another, rather unphysical problem.  

The physical reason why the gap in the density of states remains smeared
even for very long correlation lengths in the model with Gaussian phonons
resides with the fluctuations of the gap {\em amplitude} inherent in the 
model.  This smearing should not be interpreted as a presence of a large
(of order $\delta$) scattering rate.  Rather, it should be regarded as
an inhomogeneous broadening of energy levels, which, being a reversible
process, can be distinguished from relaxational broadening.  To do so,
one may attempt to study the fermion lineshape using time-resolved 
spectroscopy rather than frequency-domain methods (cf.~NMR).  

Finally, I have discussed a few aspects of the newly proposed scenario 
for the behavior of electrons at ``hot spots'' in cuprate superconductors
\cite{Schmalian}.  In particular, it appears that the antiferromagnetic
fluctuations alone cannot explain the presence of a strong pseudogap 
seen by local probes of the density of states, such as tunneling spectroscopy
and NMR.  As has been noted before \cite{not tied}, pairing fluctuations 
seem to be a necessary ingredient to explain the pseudogap at low 
frequencies.  

\section*{Acknowledgments}

I thank H.~C.~Ren, M.~V.~Sadovskii and J.~Schmalian for helpful 
discussions.  This work has been supported by the US Department 
of Energy and NEDO (Japan).

\end{document}